# Limits of Educational Soft "GeoGebra" in a Critical Constructive Review

**Valerian Antohe,**
**Faculty of Engineering, Brăila,**
**"Dunărea de Jos" University of Galaţi, România**

**ABSTRACT**. Mathematical educational soft explore, investigating in a dynamical way, some algebraically, geometrically problems, the expected results being used to involve a lot of mathematical results. One such software soft is GeoGebra. The software is **free** and **multi-platform** dynamic mathematics software for learning and teaching, awards in Europe and the USA. This paper describes some critical but constructive investigation using the platform for graph functions and dynamic geometry.

## Introduction

GeoGebra represents dynamic mathematical software for all levels of education that joins arithmetic, geometry, algebra and calculus. On the one hand, GeoGebra is an interactive geometry system (freely available from www.geogebra.org). We can do constructions with points, vectors, segments, lines, conic sections as well as functions and change them dynamically afterwards. On the other hand, equations and coordinates can be entered directly. Thus, GeoGebra has the ability to deal with variables for numbers, vectors and points, finds derivatives and integrals of functions.

## 1. Graphic lecture of real functions

Integrating Dynamic Learning Systems, such as Geometer's Sketchpad and Geogebra, and other cognitive tools, such as spreadsheets and Computer Algebra Systems, into math contests and developing online environments for e-contests may provide more opportunities to the students in terms of accessibility, promoting their interests towards mathematics, and encouraging them to advance their own cognitive abilities, [AGO02].

Many real functions can be a difficult for students in an analytical classic investigation. For example, the functions f,g:**R**-{0}→**R**, f(x)=$x \sin\left(\frac{1}{x}\right)$ or g(x)=$\frac{1}{x} \sin x$ could produce some difficulties when investigating them in a neighborhood of zero. GeoGebra program can produce real revelations as we see in the figure of the application, (fig.1 and fig.2).

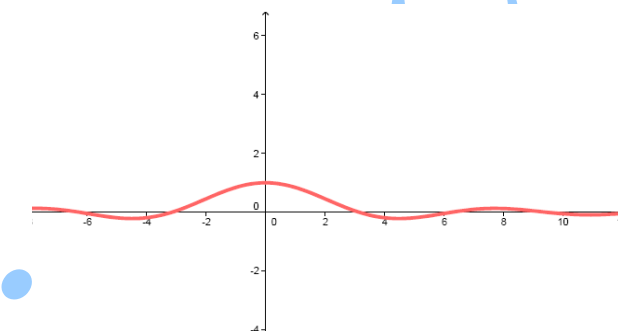

***Fig. 1***. g:**R**-{0}→**R**, g(x)=$\frac{1}{x} \sin x$

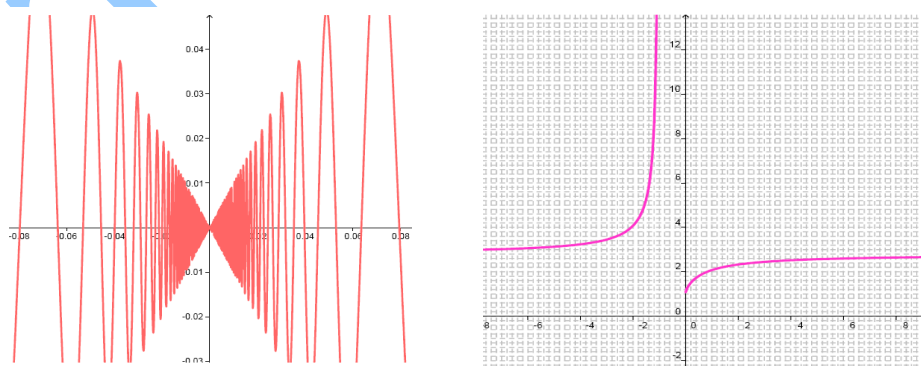

***Fig. 2***. f:**R**-{0}→**R**, f(x)=$x \sin\left(\frac{1}{x}\right)$, h: **R**-{0}→**R**, **h(x)=$\left(1+\frac{1}{x}\right)^x$**

48

In regard to the content and format of the study, we spent relatively more time to develop relevant problems for this study in contrast to developing paper-and-pencil problems. As opposed to our challenge, students showed no hesitation during the problem-solving sessions, and declared no concern when we asked about their experiences. But, this is not all! Maths can not be of real succes if we do not prove what is happening around zero! And these are the limits of the e-learning useing the computer, (fig.2).

## 2. GeoGebra in geometry

The following problem was discovered by Gheorghe Tztzeica (a Romanian mathematician). The mathematical and cultural work of the Romanian geometer Gheorghe Tzitzeica is a great one, because of its importance, its originality but also due to its dimensions: more than 200 printed papers and books with numerous editions (the problem book in geometry has more than 12 editions).

Once upon a time, while he was playing by drawing circles with „a coin" on a piece of paper. I don't believe that he was really using a golden coin, I have just made this up and the entire world can do such hipotesa (in fact he has just a five old lei currency). But the problem is beautiful enough to deserve a "golden" name. The exact mathematical statement of the problem is as follows: Consider three congruent circles of centers A,B,C, that have a common point O. Let P, M, N, be the other intersections of each pair of circles, as in the figure 3.

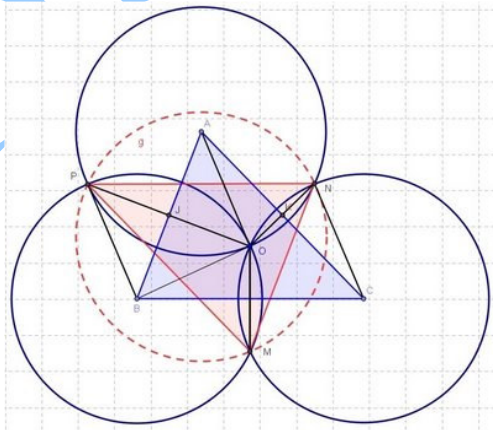

***Fig. 3***. *The 5-dollar coin problem discovered in 1908 by the Romanian Mathematician  G. Tzitzeica.*

Show that the circle circumscribed to the triangle MNP is congruent to the three initial circles. (In other words, we must show that the radius of the circumcircle of MNP is the same as the radiuses of the initial circles.)
We try to solve some of the steps indicated below using GeoGebra and we'll eventually arrive to a complete solution to the problem:

**Step 1**. In general, we prove that if two congruent circles intersect in two points, then the points of intersection and the centers of the two circles form a rhombus, (fig.4).

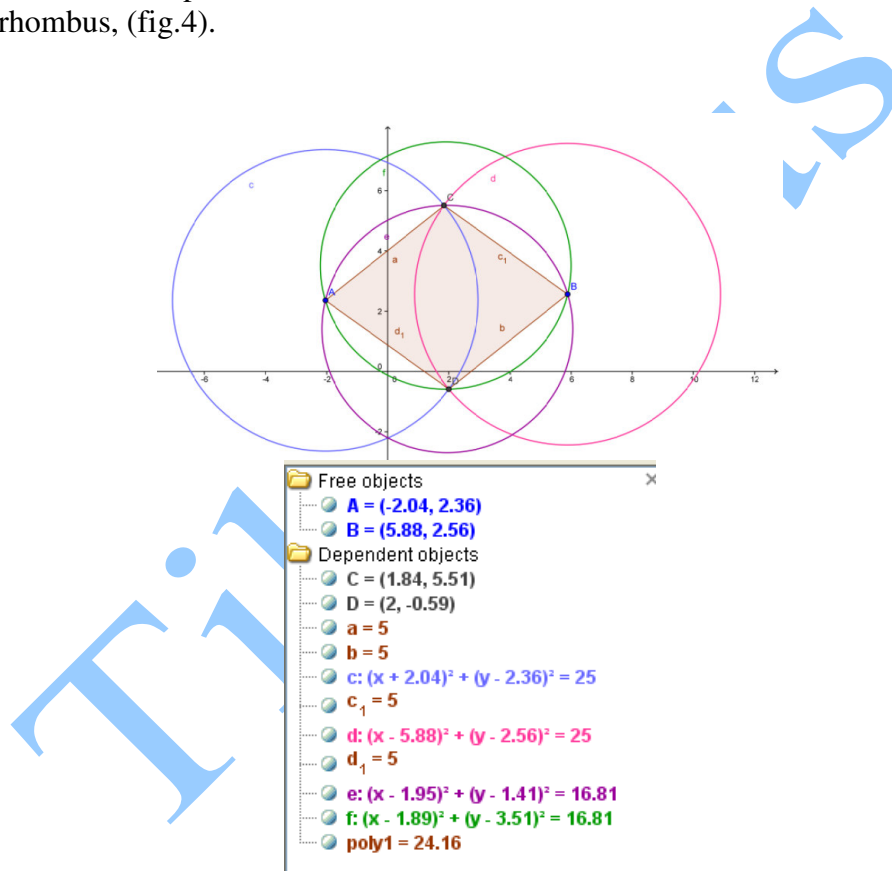

***Fig. 4***. *A dynamic geometric structure with two free points A and B*

As we see, all of a, b, c and d are five units length. In the same way, circles e and f have the same razes in their analytical equations.

**Step 2**. We'll try to arrange all of hypothesis like in Tzitzeica problem, (fig.5).

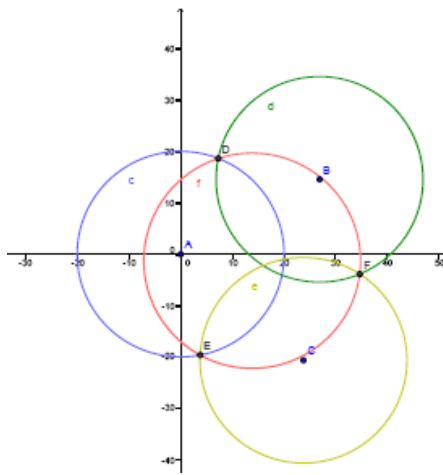

***Fig. 5***. *We must have four circles like in Tzitzeica hypotheses problem*

The protocol of the construction using GeoGebra could not construct three circles as in the Tzitzeica hipothesis, (fig.6). So, our construction will be developed as follow: first two circles, after this a segment with the length as the rases length and in the end the third circle. We'll do that because the program could not construct a circle if and only if we have three parameters: the center, the rase and a fixed point the circle cross.

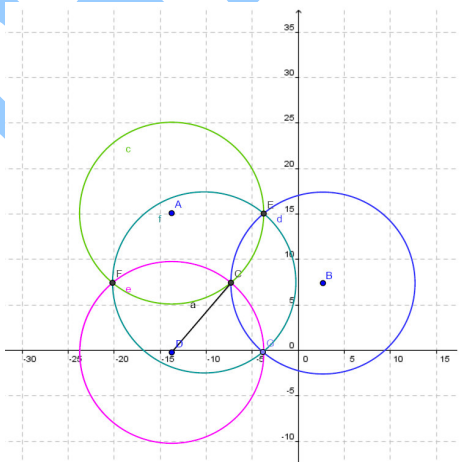

***Fig. 6***. *First try to construct three circles all with a single common point*

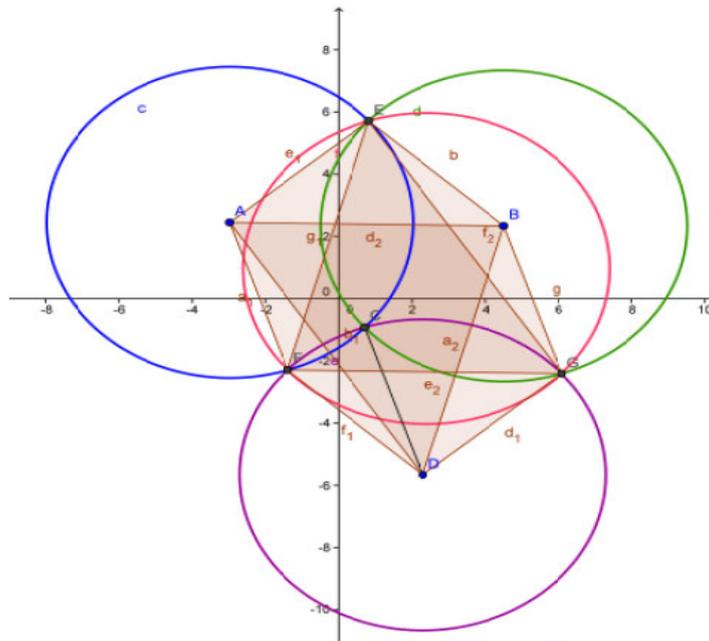

***Fig. 7***. *The most appropriated construction realized.*

As we can see in the construction protocol, although we try to obtain the razes of circles c, d, e and f of 10 units, the deviation is between 100-99.19.

The third attempt is better than the previous one but, not enough! Although we'll have to use some of the rhombuses from the figure to see that some nice parallelograms are formed and obtain the equality of the sides, a lot of our expectation don't match with exact mathematical demonstrations.

Seeing all these, we obtained the rases of circle f in the analitical equation about $r^2=24,00$. and other predictable results such are the areas of ABC triungle or GFE triungle about 30,05 units, (fig.7).

**Conclusion**

We agree that GeoGebra could be an efficient platform for e-learning. The soft is free and the comunity which works together developes a lot of exemples for arithmetic, geometry, algebra and calculus adventure, [HH02].

Even though such important theorem will find a special place in the platform, all the users must agree that the investigation is a constructive critical way, not so often restricted.

Using GeoGebra students can "see" abstract concept, students can make connection and discover mathematics. The ability to assess student solutions electronically may promote students' interests towards mathematics and advance students' cognitive abilities, [KD08]. In addition, performing e-contests in online environments can allow more students to access and benefit from math contests.